# 1. Title Page

Graph identification of proteins in tomograms (GRIP-Tomo)

August George[1,2], Doo Nam Kim[3], Trevor Moser[1], Ian T. Gildea[1], James E. Evans[*1,4], Margaret S. Cheung[*1,5]

[1] Environmental Molecular Sciences Laboratory, Pacific Northwest National Laboratory, Richland, WA, 99354, USA

[2] Department of Biomedical Engineering, Oregon Health & Science University, Portland, OR, 97239, USA

[3] Biological Science Division, Pacific Northwest National Laboratory, Richland, WA, 99354, USA

[4] School of Biological Sciences, Washington State University, Pullman, WA, 99164, USA

[5] Department of Physics, University of Washington, Seattle, WA, 98195 USA

* Co-Corresponding authors:

Margaret S. Cheung; Address: 1100 Dexter Ave N, Seattle, WA 98109; Phone: (509) 371-6486; email: margaret.cheung@pnnl.gov.

James E. Evans; Address: 3335 Innovation Blvd, Richland, WA 99354; Phone: (509) 371-6709; email: james.evans@pnnl.gov.

Running title: GRIP-Tomo

Total number of manuscript pages: 35

Total number of supplementary material pages, tables, and figures: 9 pages

Supplementary information:

Contains additional figures, including: the ribbon and graph network representations of Ankyrin repeats (Figure S1), a four helix bundle (Figure S2), hemoglobin (Figure S3), and apoferritin (Figure S4), an overview of fundamental graph properties and features (Figure S5), an overview of the procedure to convert 3D protein structures in graphs (Figure S6).

## 2. Abstract

In this study, we present a method of pattern mining based on network theory that enables the identification of protein structures or complexes from synthetic volume densities, without the knowledge of predefined templates or human biases for refinement. We hypothesized that the topological connectivity of protein structures is invariant, and they are distinctive for the purpose of protein identification from distorted data presented in volume densities. Three-dimensional densities of a protein or a complex from simulated tomographic volumes were transformed into mathematical graphs as observables. We systematically introduced data distortion or defects such as missing fullness of data, the tumbling effect, and the missing wedge effect into the simulated volumes, and varied the distance cutoffs in pixels to capture the varying connectivity between the density cluster centroids in the presence of defects. A similarity score between the graphs from the simulated volumes and the graphs transformed from the physical protein structures in point data was calculated by comparing their network theory order parameters including node degrees, betweenness centrality, and graph densities. By capturing the essential topological features defining the heterogeneous morphologies of a network, we were able to accurately identify proteins and homo-multimeric complexes from ten topologically distinctive samples without realistic noise added. Our approach empowers future developments of tomogram processing by providing pattern mining with interpretability, to enable the classification of single-domain protein native topologies as well as distinct single-domain proteins from multimeric complexes within noisy volumes.

**Statement:** The objective is to develop a new algorithm that captures topologically invariant features from densities in volumes as learning features of measuring the morphology and the hierarchy of the data sets to identify embedded protein structures.

## 3. Abbreviations and symbols

TDA, topological data analysis

PDB, protein databank

## 4. Introduction

Tomography is an imaging method which reconstructs 3D volumes of target structures from 2D cross-sections and has numerous applications for scientific discovery. Common scanning approaches for biological specimens include positron emission tomography[1] (PET), x-ray computed tomography (XCT) [2], cryogenic electron tomography (cryo-ET) [3] and atom probe tomography[4, 5] (APT). In general, most image-based tomography approaches rely on back projecting a series of two-dimensional images of the same sample at different tilt angles tilted to generate a three-dimensional volume called a tomogram. The quality of the tomograms and ease of interpretation are highly dependent on the quality of the input two-dimensional tilt images which can suffer from low contrast (i.e. low signal-to-noise ratio, SNR), data contamination, data deformation or data loss during an acquisition by an instrument. There is an urgent need for new ways to improve the interpretability of pattern mining from the sparse volume densities to overcome these experimental and computational bottlenecks - especially for understanding the structure of protein molecules in their native environment. [3, 6]

Foremost, the complex structures of the sparse data can be portraited by concerning only the connectivity of components in space using persistent[7] homology (PH), or topological data analysis[8] (TDA). These methods quantify the topologically invariant features from the data sets by simplifying its persistent attributes. When these features are interpreted with chemically known compounds or complexes, they provide the "topological fingerprints" that can be used for better prediction of protein-ligand binding properties [9] or for identifying intermediate states in biomolecular folding systems[10] . By transforming the point data in protein structures [11, 12] [13-15] into mathematical graphs of networks, the methods of network theory[16] [12, 17] are useful for further revealing the layout of a hierarchical network representing topologically complex biomolecular structures, such as the native fold of proteins, or spatial arrangement of multimers [16, 18].

Here, we developed a new method by applying the topological data analysis (TDA) to detect topologically invariant features between spatially proximal amino acids in volume densities, and leverage the network theory order parameters that offer physically meaningful interpretations of known protein structures from the protein databank[19]. Our method clustered the lumps and voids in the volume densities into a graph representation without direct reference to any chemical bond between them, as the local connectivity between them is not always available in a tomogram. We hypothesized that the volume densities in voxels, albeit sparse, capture essential topological patterns and structural hierarchies of an embedded protein structure.

For this initial work, we demonstrate that our new approach of using graph properties, Graph Identification of Proteins in Tomograms (GRIP-Tomo), identifies protein structures with high accuracy from simulated tomographic volumes lacking noise. To accelerate the identification of protein structures in tomograms, here we introduce a pipeline to test simulated samples with common defects that present challenges for the precise identification of protein structures within noisy volume density data. Our work here demonstrates that the foundation of GRIP-Tomo is compatible with these defects such that future tuning and development can be performed to match the specific typical noise levels from different tomographic methods.[20-22]

## 5. Results

### 5.1 Network theory order parameters accurately capture topological features of structurally complex proteins and their assemblies with a proper connection

We converted a protein backbone comprised of only alpha carbons into a mathematical graph such that the spatial relationship of residues, regardless that they are covalently bonded or not, was presented by an adjacency matrix whose connectivity was determined with $d_{cut}$ (**Eqn 1**). The graphs generated from the protein backbone are a coarse-grained model which enable faster computation of network theory order parameters while retaining the essential protein topology.

To determine an appropriate length for $d_{cut}$, values ranging from 6 Å to 10 Å with an increment of 1 Å were used to generate graphs from a set of five proteins of varying size and complexity. They include two single domain proteins, a beta barrel (PDB ID: 4RLC) in **Figure 1** and ankyrin repeats (PDB ID: 4LRY) in **Figure S1**, and three multimeric protein complexes, dimer of dimer (PDB ID: 3VJF) in **Figure S2**, tetramer hemoglobin (PDB ID: 1A3N) in **Figure S3**, and a 24-mer apoferritin complex (PDB ID: 2W0O) in **Figure 2** and **Figure S4**.

We used the network order parameter *Betweenness Centrality* (BC) – the number of times that a node appears on the shortest path between two other nodes – as a representative feature to determine the structurally important alpha carbons by presenting a protein structure into a network[23, 24]. The amount of information that passes through a node based on its location within the network is shown by this parameter. Therefore, it indicates structural importance within the protein, and nodes with high BC are more likely to interact with other nodes with high betweenness centrality[25]. The graph nodes and their respective amino acid residues on the protein structure were labeled with ChimeraX[26] and inspected for topological distinctions across different proteins and $d_{cut}$ values with Gephi[27]. The target cutoff distance for determining a connection should capture pairs or groupings of nodes on the structure that have high BC, signifying the spatial importance of contact formation[28].

Using this heuristic approach, we found that for single domain proteins such as beta barrel (**Figure 1**) and ankyrin repeats (**Figure S1**) that contain long-range contacts in their sequence space but are proximal in real space, $d_{cut}= 9$ Å is best to capture them. Also, we found that the connections between nodes from graphs with $d_{cut,} = 8$ Å reveal unique topological structures at the junction or interface of multiple units. As an example, apoferritin (**Figure 2a**) has a region of high BC nodes (**Figure 2b**) on a graph at $d_{cut,} = 8$ Å that captures a vertex of the octahedral structure (**Figure 2c**). A $d_{cut,} = 8$ Å also captures essential residues at the interface of multiple units in dimer of dimer (**Figure S2**) and the tetramer hemoglobin (**Figure S3**).

Other $d_{cut,}$ cutoff distances we tested were not as well suited for identifying the structures from volume densities. $D_{cut} =6$ Å and $d_{cut} =10$ Å did not show sufficient cross-chain interaction. In addition, $d_{cut} =6$ and 7 Å produced very sparse graph layouts, while $d_{cut} =10$ Å created very dense graph layouts which suggests a poor representation of the underlying 3D backbone structure. Based on the proteins we tested, we found that $d_{cut} =8$ Å is the likely best choice of cutoff distance to define a connection between alpha carbons for identifying multimers, and $d_{cut} =9$ Å is best for identifying features in monomers. Our findings suggest that the connectivity of the graph network, as determined by the pairwise cutoff threshold $d_{cut}$, can capture different levels of hierarchical structure of the target protein, and that 8 and 9 Å cutoffs are well suited for monomer and multimeric modeling, respectively. In the following sections, we created graphs using both $d_{cut} = 8$ Å and $d_{cut} = 9$ Å for identifying protein structures from volume densities.

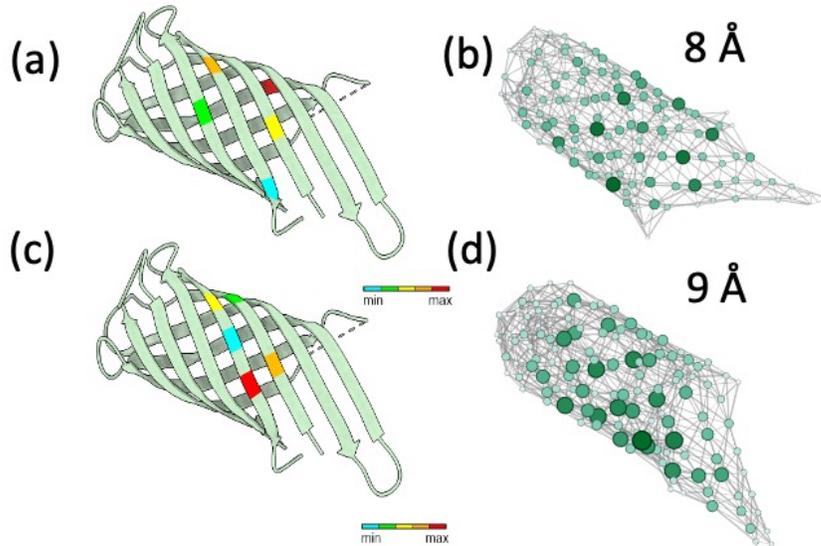

**Figure 1:** (a, c) A beta-barrel domain (4RLC). (b) and (d) are the graph network representation of the beta-barrel proteins with $d_{cut}= 8$ Å and $d_{cut}= 9$ Å, respectively. Each graph (b, d) was visualized by Gephi with the Yifan Hu force-directed layout algorithm[29]. The nodes were presented in the shades of green as well as size by betweenness (BC). The nodes with the highest BC on the graphs from (b, d) were interpreted by mapping the nodes to the amino acids on the protein structures from (a, c). We color coded these amino acids accordingly. The protein structures were visualized using Chimera.

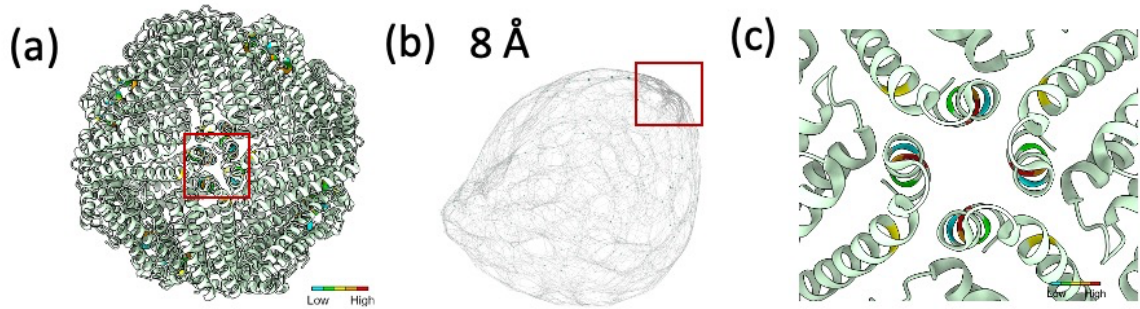

**Figure 2:** (a) The ribbon representation of the apoferritin complex structure, (b) the graph network representation of (a) with $d_{cut}$= 8 Å using the Yifan Hu force-directed layout algorithm[29]. The red box showed a cluster of nodes with highest betweenness (BC), which represents a vertex of the octahedral structure. (c) A close-up view of the vertex where residues at the most cross-unit interaction are clustered and color coded with high BC. **Figure S4** shows the graphical representations of apoferritin complex at $d_{cut}$= 7, 8, and 9 Å.

**5.2 GRIP-Tomo accurately captured the fullness of data using only the topology of a protein backbone**

Our first attempt to test the efficacy of GRIP-Tomo was to identify protein structures with only the connection of alpha carbons (Control Graphs) in a graph from the volume densities simulated with all-atom protein models (Observable Graphs) in order to address whether the GRIP-Tomo can capture the fullness of data. This set of simulated tomograms is the basis of every tested sample with minimal data deformation. (Please see the schematic workflow in **Figure 6** for the definition of a Control Graph). When full protein structures (in PDB file format) are converted into point cloud volume densities, many structural features are approximated even with high resolution. We hypothesized that GRIP-Tomo can use coarse-grain protein backbone models (from simulated tomogram data generated from full atom protein structures) to identify target proteins. We compared the similarity of ten proteins (**Table 1**) with $d_{cut}$=8Å and $d_{cut}$=9Å in **Figure 3a** and **3b**, respectively. The densities of a tomographic volume were constructed from all-atom protein structures with back projections using a full range of rotational angles from 90 to -90 degree in a 0.5-degree increment. These densities were thresholded as determined by empirical trials and converted into graphs using GRIP-Tomo. We then computed the Similarity Score, $\chi$, across all proteins using **Eqn 2**.

Along the diagonal in **Figure 3** there is a high similarity ($\chi$ >0.8) between the Control Graphs and Observable Graphs, signifying high fidelity in identifying the correct protein structures using GRIP-Tomo. Note that the graphs with $d_{cut}$=9Å in **Figure 3b** achieved 100% correct classification rate between the selected set of graph features for both single domain proteins and multimeric complexes. The graphs with $d_{cut}$=8Å correctly identified all except the single domain protein of a collapsed calmodulin (1PRW) with beta barrel (4RLC) in the red and yellow boxes in **Figure 3a.** The above results indicate that representative graph networks from the coarse-grained representation of protein structures (e.g. protein backbone) are sufficient to classify and identify proteins from point-cloud data generated using full protein structures.

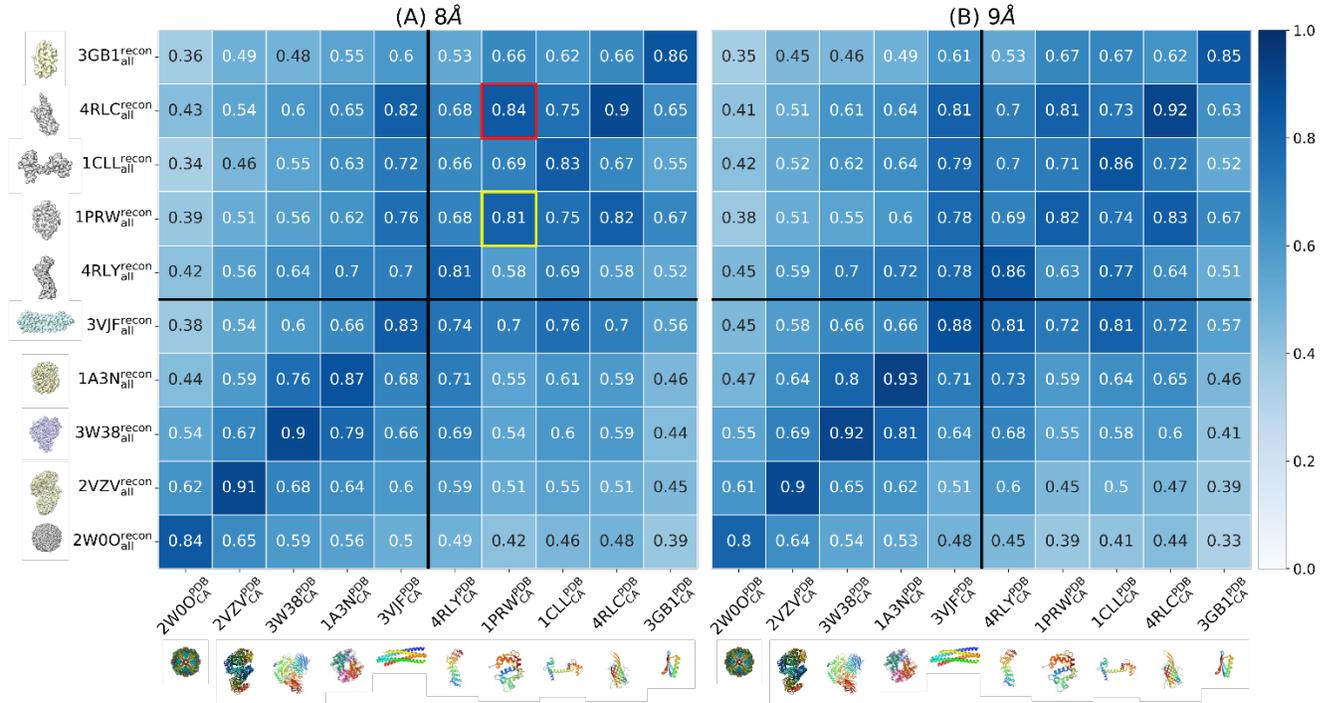

**Figure 3:** The similarity scores ($\chi$) by comparing the Observable Graphs (y-axis) against the Control Graphs (x-axis) for 10 protein structures or complexes for testing whether GRIP-Tomo can classify targets even with missing data fullness. The Observable Graphs (please see **Figure 6** for definitions) were the mathematical representations of the simulated density volumes from all-atom structures from the protein databank (PDB). The densities of a sub-tomogram volume were constructed with back projections using a full range of rotational angles from 90 to -90 degree in an 0.5 degree increment. The connections between centroids from volume densities in Observable Graphs were determined at d'$_{cut}$=8Å for (a) and d'$_{cut}$= 9Å for (b), respectively. The connections between only alpha carbons along a protein structure from PDB in the Control Graphs were determined at d$_{cut}$=8Å for (a) and d$_{cut}$= 9Å for (b). False positive classifications are shown in red boxes, with their corresponding true positive classification shown in yellow boxes.

**5.3 GRIP-Tomo minimizes the defects from tumbling due to random rotations of samples**

Our second attempt to test the efficacy of GRIP-Tomo was to identify protein structures with only the connection of alpha carbons (Control Graphs) in a graph from the volume densities (Observable graphs) with data defects due to tumbled proteins. When taking a snapshot of a native cell environment via tomography, each protein molecule is randomly tumbled except in some limited cases of organelle or location- based preferential orientation. Thus, when trying to interpret tomograms one needs to account for rotational tumbling.

Because the volume densities are deep with information from connected centroids made of atoms, we hypothesized that the application of network order parameters will advance the protein identification by revealing distinct hierarchies in the structural feature of a protein, avoiding the data deformation due to random orientations of samples. We emulated this tumbling effect in the volume densities by randomly rotating a protein in real space. The densities of a volume were constructed with back projections using a full range of rotational angles from 90 to -90 degree in an 0.5-degree increment. We then applied GRIP-Tomo (with same threshold as the one used in previous section) for identification against the Control Graph and computed the similarity score (**Figure 4**). Along the diagonal in **Figure 4** there is a relatively high similarity ($\chi > 0.78$) between the Control Graphs and Observable Graphs. Using $d_{cut} = 9$Å (**Figure 4b**) we achieved 100% correct classification rate, while the graphs with $d_{cut} = 8$Å in **Figure 4a** falsely identified a single-domain protein (collapsed calmodulin, 1PRW) and a complex (dimer of dimer, 3VJF). Our results demonstrate that GRIP-Tomo is able to achieve a high level of accuracy in identifying target structures that have undergone random tumbling, successfully disentangling the rotational degrees of freedom from the translational degrees of freedom in a complex system.

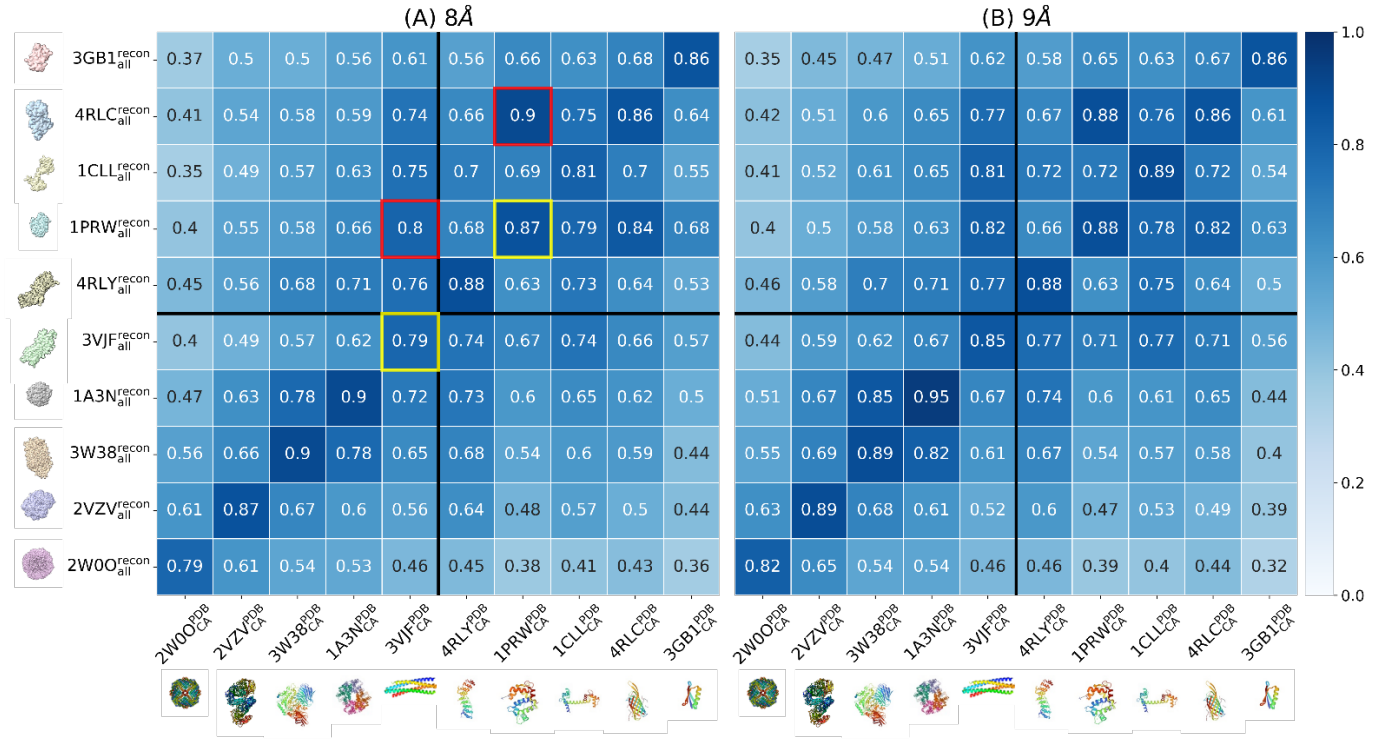

**Figure 4:** The similarity scores ($\chi$) by comparing the Observable Graphs (y-axis) against the Control Graphs (x-axis) for 10 protein structures or complexes for testing the tumbling effect. The Observable Graphs (please see **Figure 6** for definitions) were the mathematical representations of the simulated density volumes from all-atom PDB structures. In addition, the protein structures were randomly rotated in real space, mimicking the tumbling effects. The densities of a sub-tomogram volume were constructed with back projections using a full range of rotational angles from 90 to -90 degree in an 0.5 degree increment. The Observable Graphs were created at d'$_{cut}$=8Å

for (a) and d'$_{cut}$= 9Å for (b). The Control Graphs were created from the protein backbones with only alpha carbons from PDB structures at d$_{cut}$=8Å for (a) and d$_{cut}$= 9Å for (b). False positive classifications are shown in red boxes, with their corresponding true positive classification shown in yellow boxes.

## 5.4 GRIP-Tomo shows high fidelity in protein identification from the volume densities with the missing wedge effect

Our third attempt to test the efficacy of GRIP-Tomo was to identify protein structures with only the connection of alpha carbons (Control Graphs) in a graph from the volume densities (Observable graphs) with tumbled proteins and the "missing wedge effect". The missing wedge effect is a known defect that results from tomographic reconstructions generated from datasets not spanning a full 180-degree tilt wedge. Cryo-ET is one tomography method that is impacted by the missing wedge due to the limit of data acquisition typically in the range of -60° to 60° spanning a wedge of only 120 degrees. The lack of data within the missing wedge means that reconstruction in Fourier space is not only discretely populated along the given tilt angles (with zero information between each tilt angle), but that the entire space from –90 to –60 and +60 to +90 degrees also lack phase and amplitude information about the sample. The resulting defect appears as a blurring out or stretching of volume density perpendicular to the tilt axis with smaller wedges or wedges with smaller sampling (fewer total tilt images) showing more exacerbated defects. Because the network theory order parameters obey rotational invariance as shown above, we hypothesized that the topological connectivity of protein structures is invariant with respect to rotation and translation, and they are distinctive for the purpose of protein identification from distorted data presented in volume densities.

We created Observable Graphs from back projected densities r using only a partial range of rotational angles from 60 to -60 degree in a 0.5-degree increment to mimic the "missing wedge effect". Using GRIP-Tomo (with same threshold as the one in previous sections) we computed the similarity score against those from the Control Graphs in **Figure 5**. Along the diagonal in **Figure 5** there is a relatively high similarity ($\chi>0.75$) between the Control Graphs and Observable Graphs. We showed that both graphs at $d_{cut}= 8$Å (**Figure 5a**) and $d_{cut}= 9$Å (**Figure 5b**) have correctly identified 8 cases out of 10. In both cases, extended calmodulin (1CLL) is incorrectly classified as the alpha helix bundle (3VJF), and the beta sheet barrel (4RLC) is incorrectly classified as collapsed calmodulin (1PRW). Nonetheless, all of the larger proteins as well as multimers are identified correctly for this 120-degree tilt wedge (241 total tilt image) example, indicating that GRIP-Tomo can identify proteins with high-order complexities even with the missing wedge effect distorting the data.

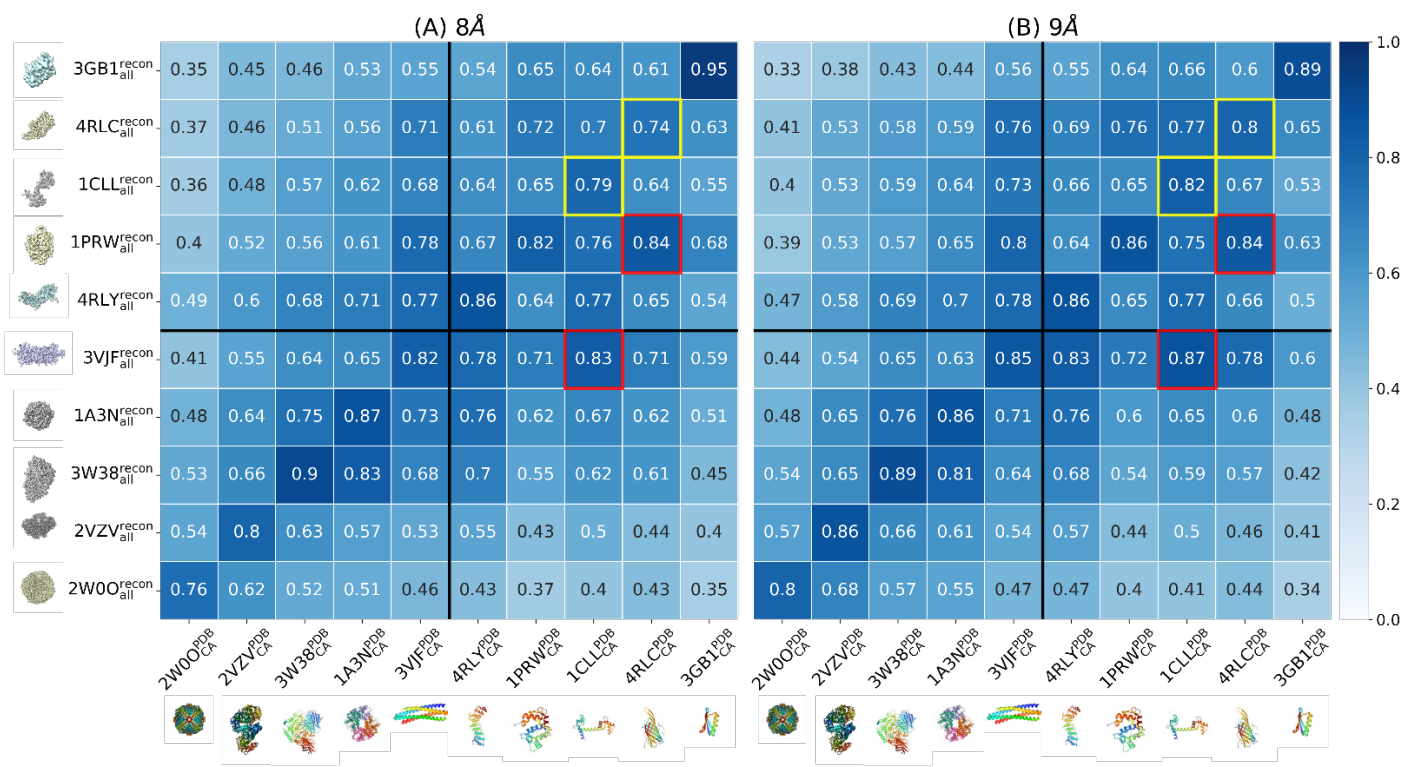

**Figure 5:** The similarity scores (χ) by comparing the Observable Graphs (y-axis) against the Control Graphs (x-axis) for 10 protein structures or complexes for testing the missing wedge effect (generated from randomly tumbled molecules). The Observable Graphs (please see **Figure 6** for definitions) were the mathematical representations of the simulated density volumes from all-atom PDB structures. The protein structures were randomly rotated in real space, mimicking the tumbling effect. In addition, the densities of the volume were constructed with back projections using a partial range of rotational angles from 60 to -60 degree in a 0.5-degree increment, mimicking the "missing wedge effect". The Observable Graphs were created at $d'_{cut}=8Å$ for (a) and $d'_{cut}= 9Å$ for (b). The Control Graphs were created from the protein backbones with only alpha carbons from PDB structures at $d_{cut}=8Å$ for (a) and $d_{cut}= 9Å$ for (b). False positive classifications are shown in red boxes, with their corresponding true positive classification shown in yellow boxes.

## 6. Discussion:

### 6.1 Graph theory order parameters capture topologically invariant connections in protein structures

Graph-based representations of structurally complex biomolecules extend the learning of chemical and molecular data by transforming the connectivity and morphology of physical structures into graphical maps for advancing discoveries in protein-ligand binding[30], protein structure-function prediction[31], and actomyosin networks[28, 32]. In these studies, the connectivity is inspired by the determination of chemically bonded and non-bonded contacts, allowing the transformation of 3D structures into 2D contact maps.

We used only alpha carbons (C$\alpha$) that represent the minimalist model of a protein for probing the connectivity of their residues in space. We transformed 3D protein structures into 2D contact maps, and further present them into mathematical graphs. C$\alpha$-only protein model has been commonly adopted in the coarse-grained molecular dynamics simulations in the past decades[33-37] for providing interpretation or prediction of experimental outcomes that would otherwise be very challenging for all-atomistic molecular dynamics simulations. Although C$\alpha$-only protein model lacks the atomistic details, it is sufficient to capture key connectivity in a native topology of a protein and accelerate scientific discoveries with reduced computational costs.

The merit of presenting 2D contact maps into mathematical graphs in our work is that the connectivity in the graphs is *topologically invariant*, and resilient to data distortion in physical space.

Another important consideration is that the graphs (and graph properties) are highly interpretable. As demonstrated in the Results section, we found that a careful selection of $d_{cut}$ and $d'_{cut}$ was necessary to capture protein structural features and connectivity. A cutoff of 8Å better captures short range interactions and protein features such as protein unit junctions, where as a cutoff of 9Å better captures longer range interactions and protein features that are distant in sequence space but relatively close in physical space. Future work will explore how to integrate networks with different connectivity (i.e. $d_{cut}$ values) in order to utilize the hierarchical information of the protein network features to improve classification and identification. Hence, GRIP-Tomo achieves a relatively high accuracy of protein identification across various data distortions in network configurations such as missing data fullness **(Figure 3)**, the tumbling effect (**Figure 4**) and the missing wedge effect (**Figure 5**) at least in the case with no noise added to the simulated volumes. These findings validate our hypothesis that the network theory order parameters are effectively invariant under the conditions studied. This further implies that these order parameters can serve as robust features for a future data-driven workflow under development.

Among the limited example of misidentified protein structures, we note that hierarchical helical structures (e.g. four helix bundle, calmodulin) prove to be challenging to correctly classify. This can likely be remedied with more hyperparameter tuning as well as a greater emphasis on global or higher order hierarchical structure such as the number of communities and assortativity[32]. Also, while the network order parameters are in theory rotationally invariant, we found some variation in the similarity scores with the proteins randomly rotated in the 3D volumes (**Figure 4**). It is possible these variations are due to specific morphologies or aspect-ratios for the particular proteins and this will be examined in future work to identify if additional constraints or dimensions are needed to fully encapsulate all samples. We also aim to develop a workflow of optimizing hyperparameters, a collection of network theory order parameters, to overcome challenges in protein identification from volume densities due to low SNR or other experimental distortions.

**6.2 GRIP-Tomo identifies proteins in a single domain and in multimeric assemblies by tuning connectivity in a network**

By varying the $d_{cut}$ that transforms 3D protein structures into 2D contact maps, we have also tuned the connectivity of a single-domain or a multimeric protein into distinct mathematical graphs. The graph properties determined at $d_{cut}$=9 Angstroms capture highly connected motifs in a tertiary structure of a protein, such as ankyrin repeats (**Figure S1**) and beta barrel (**Figure 1**), with high betweenness centrality (BC). The distance at 9 Angstroms captures long-range contacts that typically form connections across secondary structures.

In addition, by adjusting to $d_{cut}$= 8 Angstroms, the contact maps effectively capture cross-unit contacts in homo-multimeric protein complexes such as dimer of dimer, hemoglobin, and apoferritin. We were able to identify protein complexes with high accuracy from simulated volume densities with data distortion. Our future work will aim at identifying single-domain proteins from protein complexes by tuning the $d_{cut}$ in GRIP-Tomo. We note that our sample size of proteins is limited. By expanding the data set of "Control Graphs" beyond the 10 proteins we used in this work, we aim to broaden the discovery of proteins in heterogeneous samples.

**7. Conclusion**

Overall, the results indicate a relatively high accuracy (80%-100%) of the GRIP-Tomo based particle classification under a range of different data conditions evaluating the effect of missing data fullness, tumbling, and missing wedge effect without noise. We demonstrate that graph properties from protein structure networks and 3D volume densities are topologically invariant and enable rapid protein identification from point cloud/tomogram datasets. While there is more work to be done in order to improve classification and identify proteins within experimental datasets, this study serves as a proof-of-concept for the near-term application of topological data analysis to identify volumetric features from cryo-ET, PET, MRI, or CT datasets, with applications across multiple disciplines for improved object detection and pattern mining.

## 8. Methods

### 8.1 Network theory order parameters as features for protein structure networks

The graph representation $G=(V,E)$ of a snapshot of a protein structural network is defined a set of nodes $V$, which correspond to the alpha-carbon on each residue, and a set of edges E, which accounts for the relations between nodes. An edge exists between any two nodes if their two corresponding residues in the system are within a cutoff distance $d_{cut}$. We define the distance $d(v, w)$ between nodes $v$ and $w$ to be the minimum Euclidean distance among all pairs of nodes. Mathematically, the graph representation of the proteins structural network is defined as an adjacency matrix

$$\mathbf{A}_{vw} = \begin{cases} 1, & d(v,w) < d_{cut} \\ 0, & \text{otherwise} \end{cases} \quad \text{Eqn. 1}$$

where is $d_{cut}$ is an important hyperparameter governing the connectivity of the graph (see Results).

The number of vertices in the graph $G=(V,E)$ is $|V| = N_{residues}$ and the number of edges $|E|$ varies depending on the spatial distribution of the residues. Topological measures of a graph can be used as order parameters at varying graininess from a "node", a "local", to a "community" viewpoint. The network theory order parameters are sorted into three vantage points: (a) the topology of a node, which is illustrated in **Figure S5**; (b) the centrality of a node, which quantifies the connectivity or reveal the 'importance' of the node in a local network; and (c) the detection of a distinctive community in a global network. In our prior work, we have introduced several network-theory order parameters in great details as they collectively describe the hierarchical features of a network rich with biological information[28, 32]. Here, we have used a set of 12 network theory order parameters, such as number of nodes, number of edges, density, diameter, average path length, average clustering, max closeness centrality, max eigenvector centrality, max betweenness centrality, degree assortativity, max clique number, and number of communities, to capture the topological and hierarchical features of protein complexes as structural networks in a mathematical graph.

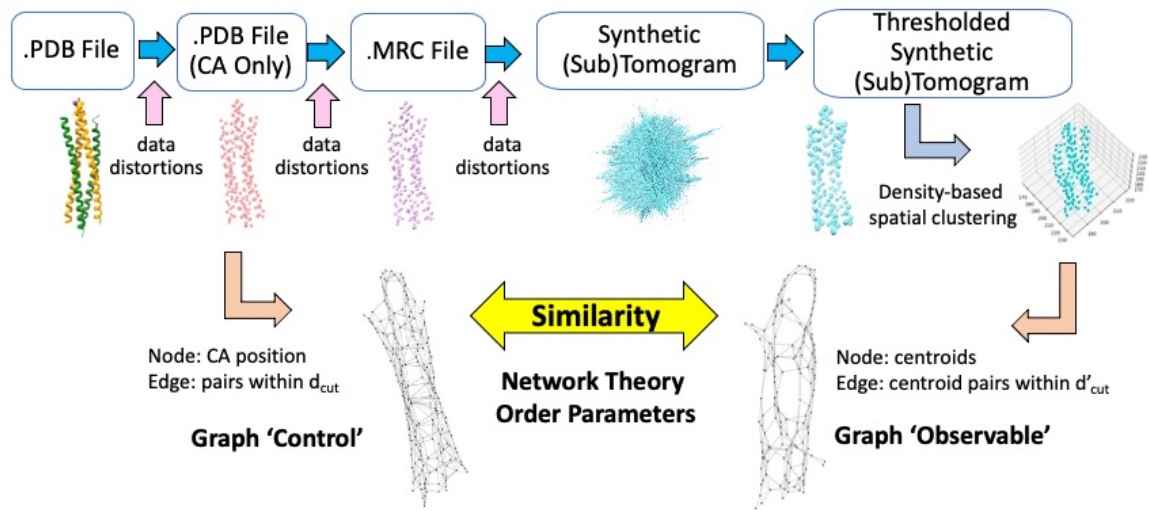

**Figure 6:** The workflow of graph identification of proteins in tomograms (GRIP-Tomo).

## 8.2. **Gr**aph **I**dentification of **P**roteins in tomograms (GRIP-Tomo)

We developed a novel data science approach based on network theory to accelerate the accurate identification of proteins and protein complexes from idealized synthetic data (i.e. point cloud densities). We created a pipeline (**Figure 6**) that classifies proteins (via a score ranking) in 3D image data based on a pool of known protein structures. This workflow is divided into four processes: 1) converting the pool of known protein structures into graphs, 2) simulating tomographic volumes of proteins with varying distortions of the data, 3) generating a graph from the tomogram, and 4) calculating a similarity score for classification. These four processes are described in detail below.

**8.2.1 Converting known protein structures into mathematical graphs**

We converted known protein structures with atomic coordinates (.PDB file) into mathematical graphs. The alpha carbon atoms were used to capture a minimal representation of the protein structure. We parsed a PDB structure file and converted it into a mathematical graph $G(V,E)$ where V is each alpha carbon in the structure. The edges are E assigned based on a pairwise Euclidean distance between nodes less than $d_{cut}$. $D_{cut}$ is determined by the user based on domain knowledge, and we used both $d_{cut}$ of 8 and 9 Angstroms for our analysis (see **Results** and **Figure S6**).

To validate our pipeline against different types of data distortions, we selected 10 proteins (see **Table 1**) with distinct secondary, ternary, quaternary structures, and sizes from the RCSB Protein Data Bank (PDB)[19] as a ground truth. The graphs generated from these known protein structures with atomistic coordinates are served as a set of Control Graphs for identifying structures from the simulated volume density data.

**Table 1:** Selected proteins for method development and validation

| Protein | PDB ID | Number of Amino Acid Residues | Stoichiometry | Structure |
|---|---|---|---|---|
| Protein G (B1 domain) | 3GB1 | 56 | Monomer | 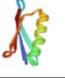 |
| b-sheet barrel | 4RLC | 136 | Monomer | 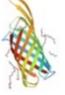 |
| Calmodulin – extended | 1CLL | 144 | Monomer | 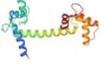 |
| Calmodulin – collapsed | 1PRW | 147 | Monomer | 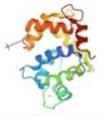 |
| 4-helix bundle | 3VJF | 175 | Dimer | 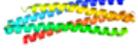 |
| Ankyrin repeats | 4RLY | 308 | Monomer | 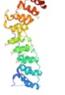 |
| Hemoglobin | 1A3N | 572 | 4-mer | 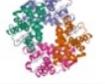 |
| Alpha-glucosidase | 3W38 | 826 | Monomer | 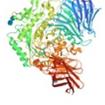 |
| Exo-chitosanase | 2VZV | 1,714 | Monomer | 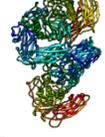 |
| Apoferritin | 2W0O | 4,250 | 24-mer | 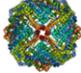 |

**8.2.2 Simulating 3D volumes from protein structures with varying controlled effects of data distortion**

We generated simulated volumes with point cloud densities to explore the applicability of GRIP-Tomo on 3D image data with added distortions (see **Figure 7**). The volumes in this work were simulated with distortions to represent the effects from 'missing data fullness, as well the experimental conditions of 'tumbling', and 'missing wedge'. The minimalist protein backbone model was converted into a which was then used to project the 3D volume into a series of 2D images at various tilt angles. These simulated 2D density images were back projected into a 3D volume to reconstruct the sample under the various conditions (see **Figure 7**).

Specifically, each protein structure in atomic coordinates (a .PDB file) was converted to 3D grids of voxels (i.e. point clouds) based on electron density using eman2[38] and saved as a common format density map (.MRC file[39]). The 3D volume (.MRC file) was projected into a series of 2D images and stored in z-stacked MRC format (.MRCS file) using relion[40] where each z-slice represented one tilted 2D density projection of the sample. Different z-stacks were used to emulate the various conditions for data fullness, tumbling or missing wedge. These simulated 2D density images were back projected into a 3D volume using IMOD[41] to reconstruct the sample under the various conditions (see **Figure 7**).

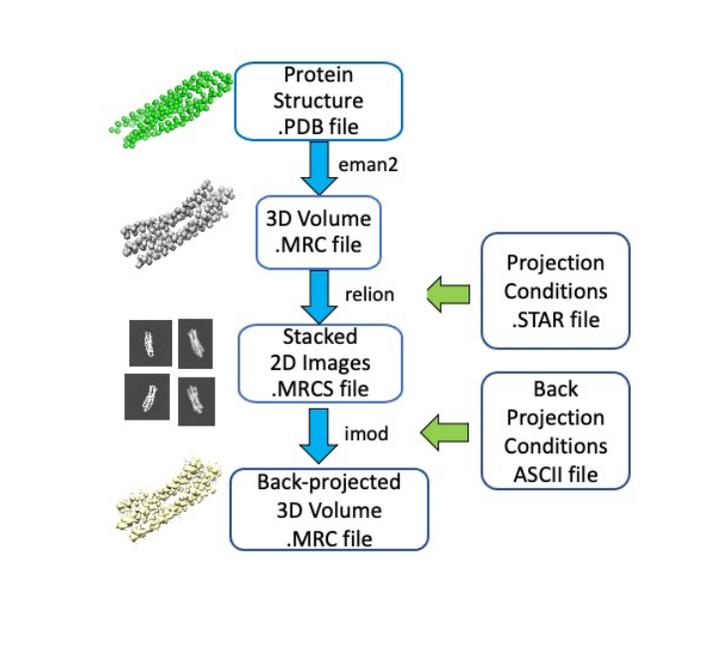

**Figure 7:** The sub-workflow of simulating a 3D density tomogram from a known protein structure.

### 8.2.3 Converting 3D densities into a mathematical graph

We first removed the low intensity values in the 3D densities in simulated volumes by applying a threshold. We used a threshold cutoff that was determined empirically using ChimeraX[26]. The remaining voxels were clustered into centroids using a density-based clustering algorithm[42] (DBSCAN). The centroid of each cluster was set as a node of the graph. Edges were assigned based on a pairwise Euclidean distance, $d'_{cut}$ of both 8 and 9 pixels (equivalent to 8 and 9 Å to match the Control Graphs). The graphs generated from these 3D densities are Observable Graphs to be compared against the Control Graphs. from **section 8.2.1**.

### 8.2.4 Measuring the similarity of the two graphs by comparing their topological features using network theory order parameters

For either the "control" graph from **section 8.2.1** or the "observable" graph from **section 8.2.3**, we first calculated the following network theory order parameters for extracting the distinct features from each mathematical graph using NetworkX[43]: number of nodes, number of edges, density, diameter, average path length, average clustering, max closeness centrality, max eigenvector centrality, max betweenness centrality, assortativity, max clique number, and number of communities.

To compute a similarity score ($\chi$) in Eqn 2, the features of each Observable Graph were compared against the features of each Control Graph. The score was calculated using the average relative similarity across all features. For each Observable Graph, if it generated the highest similarity with its respective Control Graph, the classification was determined to be true positive result.

$$c = \frac{1}{N}\sum_{i=1}^{N} w_i Z_i \text{ where, } Z_i = 1 - \frac{|X_i - Y_i|}{max(X_i, Y_i)} \qquad \textbf{Eqn 2.}$$

Here N is the total number of applied network order parameters (i.e. graph features), $w_i$ is the weight of the *i*th graph feature (set equal to one), and $X_i$ and $Y_i$ are the calculated values of the *i*th graph feature between the Observable Graph (from simulated 3D volumes) and the Control Graph (from atomic coordinates).

**Implementation**: the scripts and supporting documentation are available on GitHub at https://github.com/EMSL-Computing/grip-tomo

## 9. Supplementary material description

*Figures:* The ribbon and graph network representations of Ankyrin repeats (Figure S1), a four-helix bundle (Figure S2), hemoglobin (Figure S3), and apoferritin (Figure S4), an overview of fundamental

graph properties and features (Figure S5), an overview of the procedure to convert 3D protein structures in graphs (Figure S6).

## 10. Acknowledgement

The research was performed using resources available through Research Computing at Pacific Northwest National Laboratory (PNNL). PNNL is operated by Battelle for the U.S. Department of Energy (DOE) under Contract DE-AC05-76RL01830. A portion of the research described in this paper was conducted under the Laboratory Directed Research and Development Program at PNNL. A portion of this research was performed on a project award (DOI: https://doi.org/10.46936/intm.proj.2021.60121/60001438) from the Environmental Molecular Sciences Laboratory, a DOE Office of Science User Facility sponsored by the Biological and Environmental Research program under Contract No. DE-AC05-76RL01830. This work was supported in part by the U.S. DOE, Office of Science, Office of Workforce Development for Teachers and Scientists (WDTS) under the Community College Internship Program (CCI) to I.T.G. We also acknowledge helpful discussions with Arsam Firoozfar, Dr. Pengzhi Zhang, and Dr. Yossi Eliaz.

The authors declare that they have no conflict of interest.

## 11. Author Contributions

All authors contributed to the writing and have read and agreed to the published version of the manuscript.

**August George:** conceptualization**,** methodology, software, validation, investigation, data curation, writing – original draft, writing – review & editing, visualization

**Doo Nam Kim:** methodology, software, validation, investigation, data curation, writing – review & editing, visualization

**Trevor Moser:** methodology, software, data curation, writing – review & editing

**Ian T. Gildea:** methodology, investigation, visualization

**James E. Evans:** conceptualization**,** methodology, software, writing – review & editing, resources, funding acquisition

**Margaret S. Cheung:** conceptualization**,** methodology, writing – original draft, writing – review & editing, visualization, resources, funding acquisition, supervision, project management